\documentclass[aps,pra,floatfix,twocolumn]{revtex4-2}
\usepackage[dvips]{graphicx}
\usepackage{longtable}
\usepackage{dcolumn}
\usepackage[dvips]{graphicx}
\usepackage{bm}
\usepackage{bbm}
\usepackage{slashed}
\usepackage{times}
\usepackage{nicefrac}
\usepackage{amsmath}
\usepackage{amsfonts}
\usepackage{amssymb}
\usepackage{amsthm}
\usepackage{color}

\allowdisplaybreaks
\newcolumntype{.}{D{x}{}{-1}}
\newcolumntype{w}[1]{D{.}{.}{#1}}
\newcommand{\lbr}{\langle}
\newcommand{\rbr}{\rangle}

\newcommand{\Za}{Z\alpha}

\begin{document}

\title{Recoil corrections with finite nuclear size in hydrogenic systems}

\author{Krzysztof Pachucki}
\affiliation{Faculty of Physics, University of Warsaw,
             Pasteura 5, 02-093 Warsaw, Poland}

\author{Vojt\v{e}ch Patk\'o\v{s}}
\affiliation{Faculty of Mathematics and Physics, Charles University,  Ke Karlovu 3, 121 16 Prague
2, Czech Republic}

\author{Vladimir A. Yerokhin}
\affiliation{Max–Planck–Institut f\"ur Kernphysik, Saupfercheckweg 1, 69117 Heidelberg, Germany}

\begin{abstract}

Formulas for the combined nuclear-recoil and finite-nuclear-size effects of order $(\Za)^5$ and $(\Za)^6$ are derived without any expansion in the nuclear-charge radius $r_C$, making them applicable to both electronic and muonic atoms. The obtained results are particularly relevant for high-precision determinations of root-mean-square charge radii from muonic atom spectroscopy.
We demonstrate that calculations of the atomic isotope shift based on the widely used Breit approximation give rise to an unphysical nuclear-size contribution  that is linear in the nuclear-charge radius $r_C$ at order $(\Za)^5$. This spurious term vanishes in a full QED treatment, leaving the correct contribution quadratic in $r_C$. For electronic atoms, this quadratic term is significantly smaller than the spurious linear contribution.
\end{abstract}

\maketitle

\section{Introduction}

Finite nuclear size modifies the Coulomb potential in the vicinity of the nucleus.
Although this effect occurs in the range of just a few femtometers, much smaller
than the typical localization region of the wave function of about $10^5$ fm,
the resulting shift of energy levels is significant.
For example, the $1S$-$2S$ transition in hydrogen is affected by as much as 1 MHz, which should be
compared to
the experimental accuracy of 10~Hz \cite{parthey:11, matveev:13}
and the theoretical uncertainty of 1~kHz \cite{mohr:24:codata}.
It is thus possible to determine the proton charge radius (and, simultaneously, the Rydberg constant)
from observed hydrogen transition energies. A comparison of the extracted radius with an
independent determination from muonic hydrogen \cite{pohl:10} resulted in a long-standing
discrepancy, known as the proton radius puzzle. This discrepancy  has now been largely
resolved in favor of the muonic-hydrogen radius \cite{mohr:24:codata}.

For atoms with more than one electron, absolute charge radius determinations are not yet feasible, 
as theoretical precision has not reached the required level. However, it is possible to determine differences
of nuclear-charge radii between two isotopes of the same element.
Of particular interest is the comparison of the nuclear radius differences obtained from electronic and muonic atoms. 
This comparison is highly sensitive to nuclear-polarizability effects and provides an opportunity to test fundamental interaction theories.

This field has been developing rapidly in recent years. 
For instance, the difference in squared charge radii between the deuteron and proton, $r_C^2(d) - r_C^2(p)$, 
was found to be in perfect agreement between H-D and $\mu$H-$\mu$D determinations, 
after a meticulous evaluation of deuteron polarizability effects \cite{pachucki:24:rmp}. 
Another example is the squared charge radius difference of helium isotopes, $r_C^2(^3$He$) - r_C^2(^4$He), 
which was initially reported to disagree between electronic and muonic isotope shift measurements \cite{schuhmann:23, werf:23}. However, it was recently shown that inclusion of the
second-order hyperfine-interaction correction \cite{qi:24, pachucki:24:sechfs} resolves this discrepancy.
For heavier elements, multiple isotope shift measurements have been conducted for electronic Li, Be, and heavier atoms,
while the corresponding muonic-atom measurements are currently being pursued by Ohayon {\em et al.} \cite{ohayon:24}.

The influence of the finite nuclear size (fns) on atomic energy levels has been extensively studied,
both within the expansion over the parameter $\Za$ \cite{friar:79:ap} (where $Z$ is the nuclear charge number and $\alpha$ is
the fine-structure constant) and to all orders in $\Za$ \cite{shabaev:93:fns}.
In particularly, Friar \cite{friar:79:ap} derived the fns corrections up to order $(Z\,\alpha)^6$  for the
infinitely heavy nucleus.
His derivation and final formulas has recently been verified and simplified in Ref. \cite{pachucki:18}.

Theoretical treatment of the fns effect in the presence of a finite-mass nucleus,
however, has proven to be significantly more challenging. Addressing this effect,
Friar \cite{friar:79:ap} obtained a contribution linear in the nuclear radius $r_C$
at $(Z\,\alpha)^5\,m^3/M$ order, where $m$ is the electron mass and $M$ is the nuclear mass.
If correct, this would represent a substantial contribution, as its magnitude is comparable to that 
of the leading fns recoil effect, which is of order $(\Za)^4(m^4/M)r^2_C$. 
A similar approach to the fns recoil effect was later employed by Borie and Rinker 
in their seminal work on muonic atoms \cite{borie:82} and subsequently by Borie in Ref. \cite{borie:12}.
It was, however, shown by Shabaev \cite{shabaev:98:recground} that the linear in $r_C$ fns recoil term
disappears in the rigorous QED treatment.

In this work we derive complete formulas for the recoil fns effects at orders $(Z\,\alpha)^5$
and $(Z\,\alpha)^6$. The derivation is carried out without any expansion in the nuclear radius $r_C$.
While an expansion in $m r_C$ is justified and commonly used for electronic atoms, 
it becomes entirely inadequate for muonic atoms,  where $m_{\mu}r_C\approx 1$ (with $m_{\mu}$ the muon mass). 
Moreover, we explicitly demonstrate that the term linear in $r_C$ at order $(Z\,\alpha)^5$ is an artifact of the approximate treatment of nuclear recoil. 
In this approach, the nonrelativistic nuclear kinetic energy and Breit (magnetic) interaction are added to the Dirac equation, a method that is inconsistent with QED.
Using the heavy-particle formulation of QED \cite{pachucki:23:prl, pachucki:24:hpqed}, 
we apply the exact formula for nuclear-recoil effects with finite nuclear size and demonstrate limitations of the approximate treatment.

\section{Finite nuclear size}
The finite nuclear size leads to a shift of binding energies of atomic systems, $E_{\rm fns}$.
For a light atom we can perform the expansion of $E_{\rm fns}$ in the parameter $\Za$,
where $Z$ is the nuclear charge number and $\alpha = e^2/(4\,\pi)$ is the fine-structure constant,
\begin{align}
 E_{\rm fns} =  E^{(4)}_{\rm fns} +  E^{(5)}_{\rm fns}+  E^{(6)}_{\rm fns} + \cdots\,, \label{01}
 \end{align}
where the superscript denotes the order in $\Za$.
Each term of this expansion can be further expanded in the mass ratio $m/M$,
where $m$ is the mass of the orbiting particle (electron or muon) and $M$ is the nuclear mass,
\begin{align}
E^{(n)}_{\rm fns} = E^{(n,0)}_{\rm fns} + E^{(n,1)}_{\rm fns} +\cdots\,, \label{02}
\end{align}
where the second superscript  denotes the order in the mass ratio $m/M$.

The leading-order nuclear contribution is of order $(\Za)^4$ and given by a simple formula
\begin{equation}
  E^{(4)}_{\rm fns} =  \frac{2\,\pi}{3}\,\Za\,\phi^2(0)\,r_C^2\,, \label{03}
\end{equation}
where $\phi(0)$ is the nonrelativistic electron (muon) wave function at the position of the nucleus,
$r_C$ is the root-mean-square charge radius of the nucleus,
\begin{align}
 r_C^2 =  \int d^3 r\, r^2\,\rho(\vec r)\,, \label{04}
 \end{align}
and $\rho(\vec r)$ is the nuclear charge distribution. Equation (\ref{03}) includes the exact dependence on
the finite nuclear mass $M$ through
\begin{align}
\phi^2(0) = \mu^3\,\frac{(\Za)^3}{\pi n^3}\,, \label{05}
\end{align}
where the reduced mass $\mu = mM/(m+M)$.

The next-to-leading fns correction is of order $(\Za)^5$. In the nonrecoil limit it
was obtained by Friar \cite{friar:79:ap}, with the result
\begin{align}
  E^{(5,0)}_{\rm fns} = -\frac{\pi}{3}\,\phi^2(0)\,(\Za)^2\,m\,r_F^3\,,\label{06}
\end{align}
where
\begin{align}
r_F^3 =  \int d^3r_1\int d^3r_2\,\rho(r_1)\,\rho(r_2)\,|\vec r_1-\vec r_2|^3\,. \label{07}
\end{align}
The finite nuclear mass (or, nuclear recoil) effects in Eq.~(\ref{06}) can be partially included in $\phi^2(0)$
through the reduced mass $\mu$.
However, there are further nuclear recoil corrections of order $(\Za)^5$ \cite{pachucki:23:prl},
which will be addressed in Sec.~\ref{sec:Za5recoil}.

\section{Nonrecoil correction of order $\bm{(Z\,\alpha)^6}$ for $\bm{nS}$ states}
In this section we recalculate the fns correction of order $(\Za)^6$ in the nonrecoil limit, $E^{(6,0)}$,
for the dipole (exponential) parametrization of the nuclear charge form factor 
\begin{align}
\rho(\vec q^{\,2}) = \frac{\Lambda^4}{(\Lambda^2 + \vec q^{\,2})^2}\,. \label{10}
\end{align}
The charge  radii are related to derivatives of $\rho(\vec q^{\,2})$ at $\vec q^{\,2}=0$, namely
\begin{align}
\rho'(0) =  -\frac{r_C^2}{6}\,, \ \
\rho''(0)= \frac{r_{CC}^4}{60}\,, \label{13}
\end{align}
where
\begin{align}
r_{CC}^4 = \int d^3r\,r^4 \rho(\vec r)\,.\label{13a}
\end{align}
In the exponential parametrization, the charge radii $r_C$ and $r_{CC}$ are evaluated
analytically as
\begin{align}
r_C = \frac{2\,\sqrt{3}}{\Lambda}\,,\ \
r_{CC} = \biggl(\frac{5}{2}\biggr)^{1/4}\,r_C\,. \label{15}
\end{align}
The $E^{(6,0)}$ was originally derived by Friar \cite{friar:79:ap} and a much simplified derivation was presented in Ref. \cite{pachucki:18}.
Following this simplified derivation, we  split the fns correction for an $nS$ state into high- and low-energy parts,
\begin{align}
E^{(6,0)}_{\rm fns}(nS) = E_{H} + E_{L} \,. \label{16}
\end{align}
The high-energy part $E_H$ is given by the three-photon
scattering amplitude with momenta $p_i = (m,\vec q_i)$
\begin{widetext}
\begin{align}
  E_H =&\ -(4\,\pi\,Z\,\alpha)^3\phi^2(0)\!\int\!\frac{d^dq_1}{(2\,\pi)^d} \int \frac{d^dq_2}{(2\,\pi)^d}\,
  \frac{\rho(q_1^2)}{q_1^4}\, \frac{\rho(q_2^2)}{q_2^4}\, \frac{\rho(q_3^2)}{q_3^2}\,
       {\rm Tr}\biggl[(\not\!p_1+m)\,\gamma_0\, (\not\!p_2+m)\, \frac{(\gamma_0+I)}{4}\biggr]\,, \label{17}
\end{align}
where we use the dimensional regularization with  $d=3-2\,\epsilon$, $\vec q_3 = \vec q_1-\vec q_2$, and $\phi^2(0) = \langle\phi| \delta^d(r) |\phi\rangle$.
The above trace equals to $ 4\,m^2 +\vec q_1\cdot\vec q_2$, so we split $E_H$
\begin{align}
  E_H  =&\  E_{H1} + E_{H2}\,, \label{18}
 \end{align}
 into nonrelativistic $E_{H1}$ and relativistic $E_{H2}$ parts. Using integration formulas from Appendix \ref{APPA} we obtain
 \begin{align}
 E_{H1} =&\
4\,\pi\,(Z\,\alpha)^3\,\phi^2(0)\,4\,m^{2}\,\frac{r_C^4}{36}\,
    \biggl[\frac{1}{4\,\epsilon} + 3 + \frac{7}{128}+\frac{10}{27} + 2\,\ln(2) - \frac{3}{2}\,\ln(3) + \ln(r_{C})\biggr]\,, \label{19}
\\
E_{H2} =&\
-4\,\pi\,(Z\,\alpha)^3\,\phi^2(0)\,\frac{r_C^2}{6}\,
  \biggl[ \frac{1}{4\,\epsilon} + \frac{16}{27}-\frac{3}{16}  + 2\,\ln(2) - \frac{3}{2}\,\ln(3) + \ln(r_{C})\biggr] \,. \label{20}
\end{align}
The elimination of $1/\epsilon$ singularities will be performed in atomic units,
which in $d$ dimensions become a little more complicated.
The nonrelativistic Hamiltonian in natural units is
\begin{equation}
H = \frac{\vec{p}^{\,2}}{2\,m} -Z\,\alpha\,\bigg[\frac{1}{r}\biggr]_\epsilon, \label{21}
\end{equation}
where
\begin{align}
\bigg[\frac{1}{r}\biggr]_\epsilon =&\ \int  \frac{d^d q}{(2\,\pi)^d}\,\frac{4\,\pi}{q^2}\,e^{i\,\vec k\cdot\vec r} =  \frac{C_1}{r^{1-2\,\epsilon}}\,, \label{22}
\end{align}
with
\begin{align}
C_1 = \pi^{\epsilon-1/2}\,\Gamma(1/2-\epsilon) \,.\label{23}
\end{align}
Using coordinates in atomic units
\begin{equation}
\vec r = (m\,Z\,\alpha)^{-1/(1+2\,\epsilon)}\,\vec r_\mathrm{au}, \label{24}
\end{equation}
the Hamiltonian can be written as
\begin{align}
H =&\ m^{(1-2\,\epsilon)/(1+2\epsilon)}\,(Z\,\alpha)^{2/(1+2\,\epsilon)}\,
\biggl[\frac{\vec p_\mathrm{au}^{\,2}}{2}
-\frac{C_1}{r_\mathrm{au}^{1-2\,\epsilon}}\biggr]. \label{25}
\end{align}
If one pulls out the factor
$m^{(1-2\,\epsilon)/(1+2\epsilon)}\,(Z\,\alpha)^{2/(1+2\,\epsilon)}
\approx m\,(Z\,\alpha)^2\,(Z\,\alpha\,m)^{-4\,\epsilon}$
from $H$, then one obtains the nonrelativistic Hamiltonian in atomic units.
For the leading relativistic correction, one needs to pull out the factor
$m\,(Z\,\alpha)^4\,(Z\,\alpha\,m)^{-8\,\epsilon}$.
Similarly for $H^{(6)}$, the common factor
$m\,(Z\,\alpha)^6\,(Z\,\alpha\,m)^{-12\,\epsilon}$
is pulled out from all the terms, which corresponds to the replacement
$m\rightarrow 1$ and $Z\,\alpha \rightarrow 1$ in atomic units. This factor is also pulled out
from $E_H$, which is denoted by ${\cal E}_H$,
\begin{align}
{\cal E}_{H1} =&\
4\,\pi\,\phi^2(0)_\mathrm{au}\,\frac{m^4\,r_C^4}{36}\,
    \biggl[\frac{1}{4\,\epsilon} + 3 + \frac{7}{128}+\frac{10}{27} + 2\,\ln(2) - \frac{3}{2}\,\ln(3) + \ln(Z\,\alpha\,m\,r_{C})\biggr]\,, \label{26}
\\
{\cal E}_{H2} =&\ -4\,\pi\,\phi^2(0)_\mathrm{au}\,\frac{m^2\,r_C^2}{6}\,
  \biggl[ \frac{1}{4\,\epsilon} + \frac{16}{27}-\frac{3}{16}  + 2\,\ln(2) - \frac{3}{2}\,\ln(3) + \ln(Z\,\alpha\,m\,r_{C})\biggr] \,, \label{27}
\end{align}
where
\begin{align}
\phi^2(0) = (m\,Z\,\alpha)^{d\,(1-2\,\epsilon)}\,\phi^2(0)_\mathrm{au}\,. \label{28}
\end{align}
The low-energy part ${\cal E}_L$ is obtained from the nonrelativistic expansion of the Dirac-Coulomb Hamiltonian in atomic units
\begin{align}
{H}_D =&\ \frac{\vec p^{\,2}}{2} -\biggl[\frac{1}{r}\biggr]_\epsilon -\frac{p^4}{8} +\frac{\pi}{2}\,\delta^d(r) + \delta V + \delta^{(2)}V + \frac{1}{8}\,\nabla^2(\delta V)\,,\label{29}
\end{align}
where $\delta V$ and $\delta^{(2)} V $ are the fns corrections to the Coulomb potential, given by
\begin{align}
\delta V =&\ -m^2\,\rho'(0)\,4\,\pi\,\delta^{d}(r)  =  \frac{2\,\pi}{3}\,m^2\,r_C^2\,\delta^d(r)\,, \label{30} \\
\delta^{(2)} V =&\ \frac{1}{2}\,m^4\,\rho''(0)\, 4\,\pi\,\nabla^2\delta^{d}(r) =   \frac{\pi}{30}\, m^4\,r_{CC}^4\,\nabla^2\delta^{d}(r)\,. \label{31}
\end{align}
The ${\cal E}_L$ is split into two parts ${\cal E}_L =  {\cal E}_{L1} + {\cal E}_{L2}$, where
\begin{align}
   {\cal E}_{L1} =&\  \langle\delta V\,\frac{1}{(E-H)'}\,\delta V\rangle  + \langle\delta^{(2)} V\rangle
   \nonumber \\ =&\
m^4\,r_C^4\,\frac{4}{9\,n^3}\biggl[-\frac{1}{n} -\frac12 + \gamma -\ln\frac{n}{2}+\Psi(n) \biggr]
 -m^4\,r_C^4\, \frac{\pi}{9\,\epsilon}\,\phi^2(0)+ m^4\,r^4_{CC}\,\frac{1}{15\,n^5} \,,\label{32}\\
   {\cal E}_{L2} =&\
   2\, \bigg\langle \delta V\frac{1}{(E-H)'}\,\biggl[-\frac{p^4}{8} +\frac{\pi}{2}\,\delta^d(r) \biggr]\bigg\rangle +  \bigg\langle \frac{1}{8}\,\nabla^2(\delta V)\bigg\rangle
\nonumber \\ =&\
-m^2\,r_C^2\,\frac{2}{3\,n^3}\biggl[\frac{9}{4\,n^2}-\frac{1}{n}-\frac52 +\gamma -\ln\frac{n}{2}+\Psi(n) \biggr]
  +m^2\,r_C^2\,\frac{\pi}{6\,\epsilon}\, \phi^2(0)_\mathrm{au} \,. \label{33}
\end{align}

The complete $O(\alpha^2)$ finite nuclear size correction for an arbitrary nucleus is given by the
sum $E^{(6,0)}_{\rm fns} = E_L + E_H $. The diverging $1/\epsilon$  terms in Eqs. (\ref{26}), (\ref{27}), (\ref{32}), and (\ref{33}) cancel out in the sum 
and the result is
\begin{align}
E^{(6,0)}_{\rm fns}(nS) =&\ -(Z\,\alpha)^6\,m^3\,r_C^2\,\frac{2}{3\,n^3}\,\biggl[-\frac{5}{4} + \frac{9}{4\,n^2} -\frac{1}{n}  -\ln n+\gamma+\Psi(n)
        +\kappa_1 +\ln(m\,r_{C}\,Z\,\alpha)\biggr]
        \nonumber \\
        & + (Z\,\alpha)^6\,m^5\,r_C^4\,\frac{4}{9\,n^3}\,\biggl[1-\frac{1}{n}  - \ln n  + \gamma + \Psi(n) +\kappa_2
        + \ln(m\,r_C\,Z\,\alpha)\biggr]
        +(Z\,\alpha)^6\,m^5\,r_{CC}^4\,\frac{1}{15\,n^5}\,, \label{34}
\end{align}
where
\begin{align}
\kappa_1 =&\ \frac{16}{27} -\frac{23}{16} + \frac{3}{2}\,\ln\frac{4}{3} \approx -0.413\,384\,, \label{35}\\
\kappa_2 =&\ \frac{10}{27}+\frac{199}{128}  +\frac{3}{2}\,\ln\frac{4}{3} \approx 2.356\,581\,. \label{36}
\end{align}
This is the complete result of order $(\Za)^6$ in the nonrecoil limit.
Recoil corrections can partially be accounted for by the replacement $m\rightarrow \mu$.
However, there are further nuclear-recoil effects, which are studied in the rest of the paper.

\section{Recoil fns correction}

The nuclear recoil effect to the first order in the mass ratio was described theoretically to all orders in $\Za$
for the point-like nucleus \cite{shabaev:85,shabaev:88,pachucki:95}. The nuclear recoil with the finite-size
effect has been derived in the heavy-particle QED approach \cite{pachucki:24:hpqed}.
The derived formula is
\begin{align}
E_\mathrm{rec} =&\
\frac{i}{M} \int_s \frac{d\omega}{2\,\pi}\,  \langle \phi|D_T^j(\omega) \,G(E_D + \omega)\,D_T^j(\omega)|\phi\rangle \,, \label{37}
\end{align}
where $G(E) =\nicefrac{1}{(E-H_D)}$ is the Dirac-Coulomb Green function, $H_D$ and $E_D$ are the Dirac Hamiltonian and the Dirac energy,
respectively, and
\begin{align}
D_T^j(\omega,\vec r) =&\ -4\pi Z\alpha \, \alpha^i \, G_{T}^{ij}(\omega,\vec{r})\,, \label{39}
\\
G_{T}^{ij}(\omega,\vec{r}) =&\ \int\frac{d^3k}{(2\,\pi)^3}\, \frac{\rho(-k^2)}{k^2}\,\biggl(\delta^{ij}-\frac{k^i\,k^j}{\omega^2}\biggr)\,, \label{40}
\end{align}
where $k^2 = \omega^2-\vec{k}^2$.
The subscript $s$ at the integration sign denotes a symmetric  integration around the pole at $\omega=0$ along the Feynman
or Wick rotated contour (see Fig.~\ref{fig:1}). Because the terms with $1/\omega$ singularity can be separated from terms involving  branch cuts
starting at $\omega=0$, this symmetric integration can safely be implemented.
\begin{figure}[htb]
\includegraphics[scale=0.7]{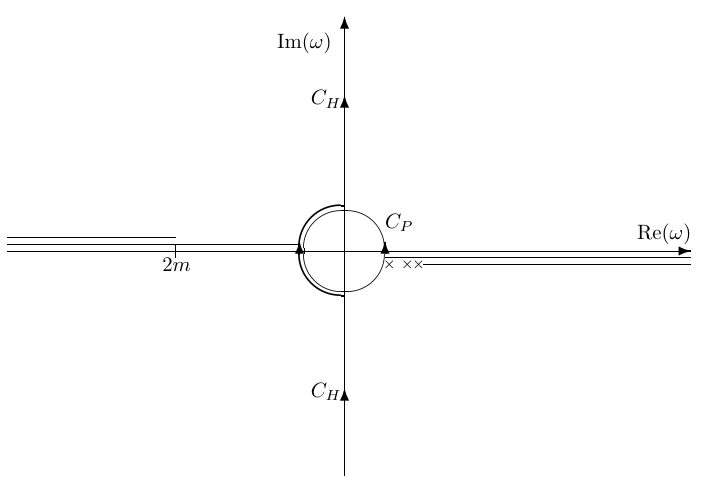}
\caption{Integration contours}
\label{fig:1}
\end{figure}

The significant difference with respect to nonrecoil corrections is in the argument of the nuclear-charge form factor $\rho$.
It is a function of $-k^2 = \vec k^{\,2} -\omega^2$ and the dependence on $\omega$ can not be neglected.
For this reason we have to assume that $\rho(-k^2)$ can be analytically continued 
to the whole complex plane, apart from the negative real axis.
Not all the charge densities used in the literature allow for analytical continuation.
The dipole parametrization is the simplest choice that has analytic continuation on the complex plane and for this reason
we use this parametrization in our work.

We now transform $E_\mathrm{rec}$ to a form that uses the photon propagator in the Coulomb gauge, using the
identity
\begin{align}
D_C^j(\omega) =&\  D_T^j(\omega)  + \frac{1}{\omega^2}\,\big[\omega+E_D-H_D\,,\,p^j(V_C)\big]\,, \label{41}
\end{align}
where
 \begin{align}
D_C^j(\omega,\vec r) =&\ -4\pi Z\alpha \, \alpha^i \, G_{C}^{ij}(\omega,\vec{r})\,, \label{42}
\\
G_C^{ij}(\omega,\vec r) =&\ \int \frac{d^3k}{(2\,\pi)^3}\,\biggl[\frac{\rho(-k^2)}{k^2}\,\biggl(\delta^{ij}-\frac{k^i\,k^j}{\omega^2}\biggr)
- \frac{k^i\,k^j}{\omega^2}\,\frac{\rho(\vec k^2)}{\vec k^2}\biggr] \,e^{i\,\vec k\cdot\vec r}\,. \label{43}
\end{align}
This Coulomb gauge form of the nuclear recoil correction is
 \begin{align}
 E_\mathrm{rec} =&\
  -\frac{i}{M} \int_s \frac{d\omega}{2\,\pi}\, \lbr \phi | \big[ p^j(V_C) - \omega\,D_C^j(\omega)\big] \,
G(E_D + \omega ) \, \big[ p^j(V_C) + \omega\,D_C^j(\omega)\big] | \phi \rbr\, \frac{1}{\omega^2}\,, \label{44}
\end{align}
where $p^j(V_C) = [p^j\,,\,V_C]$. This form is convenient to derive the leading recoil term,
which comes from the nuclear kinetic energy and the Breit interaction. We remind the reader
that the subscript "$s$" in Eq. (\ref{44}) denotes a symmetric integration around the pole at $\omega=0$, so
\begin{align}
\int_s\frac{d\omega}{2\,\pi}\,f(\omega) \,\frac{1}{\omega^2} =&\  \int\frac{d\omega}{2\,\pi}\,f(\omega) \,\frac{1}{2}\,
\biggl(\frac{1}{(\omega+\epsilon)^2} + \frac{1}{(\omega-\epsilon)^2} \biggr)\,, \label{45}
\end{align}
where the Wick rotated contour is assumed. We now split this integral in Eq. (\ref{44}) into two parts
\begin{align}
\int_s\frac{d\omega}{2\,\pi}\,f(\omega) \,\frac{1}{\omega^2} =&\
 \int\frac{d\omega}{2\,\pi}\,f(\omega) \,\frac{1}{2}\, \biggl(\frac{1}{(\omega+\epsilon)^2} - \frac{1}{(\omega-\epsilon)^2} \biggr)
 +  \int \frac{d\omega}{2\,\pi}\,f(\omega) \, \frac{1}{(\omega-\epsilon)^2}\,. \label{46}
\end{align}
The first term in this equation yields the recoil correction in the Breit approximation; it
will be referred to as the pole part in the following. It is calculated as
\begin{align}
E_\mathrm{pole} =&\
 -\frac{i}{M} \int \frac{d\omega}{2\,\pi}\, \lbr \phi | \big[ p^j(V_C) - \omega\,D_C^j(\omega)\big] \,
\frac{1}{E_D+\omega-H_D} \, \big[ p^j(V_C) + \omega\,D_C^j(\omega)\big] | \phi \rbr\, \frac{1}{2}\Biggl(\frac{1}{(\omega+\epsilon)^2} - \frac{1}{(\omega-\epsilon)^2}\biggr)
\nonumber \\ =&\
E_\mathrm{pole1} + E_\mathrm{pole2}\,. \label{47}
\end{align}
The first term here is evaluated as
\begin{align}
E_\mathrm{pole1} =&\ -\frac{i}{M} \int \frac{d\omega}{2\,\pi}\, \lbr \phi | p^j(V_C)  \,
\frac{1}{E_D + \omega -H_D} \,p^j(V_C) | \phi \rbr\, \frac{1}{2}\,\biggl(\frac{1}{(\omega+\epsilon)^2} - \frac{1}{(\omega-\epsilon)^2}\biggr)
\nonumber \\ =&\
-\frac{1}{2\,M}\,\lbr \phi | p^j(V_C)  \,
\frac{1}{(E_D - H_D)^2} \,p^j(V_C) | \phi \rbr
\nonumber \\ =&\
\lbr \phi | \frac{p^2}{2\,M} | \phi \rbr\,. \label{48}
\end{align}
The second term is
\begin{align}
E_\mathrm{pole2} =&\
-\frac{i}{2\,M} \int \frac{d\omega}{2\,\pi}\, \lbr \phi | p^j(V_C) \,G(E_D + \omega ) \, D_C^j(\omega) | \phi \rbr\,
\biggl(\frac{1}{\omega+\epsilon} - \frac{1}{\omega-\epsilon}\biggr) + \mathrm{h.c.}
\nonumber \\ =&\
\frac{1}{2\,M}\, \lbr \phi | p^j(V_C) \,\frac{1}{E_D-H_D} \, D_C^j(0) | \phi \rbr
-\frac{1}{2\,M} \, \lbr \phi | D_C^j(0) \,\frac{1}{E_D-H_D} \, p^j(V_C)  | \phi \rbr
\nonumber \\ =&\
-\frac{1}{2\,M}\, \lbr \phi | \biggl\{p^j\,,\, \frac{Z\alpha}{2}\,\biggl[\frac{\delta^{ij}}{r} + \frac{r^i\,r^j}{r^3}\biggr]_\rho \alpha^i \biggr\} | \phi \rbr\,, \label{49}
\end{align}
where
\begin{align}
\frac{1}{2}\,\biggl[\frac{\delta^{ij}}{r} + \frac{r^i\,r^j}{r^3}\biggr]_\rho =&\
4\,\pi\,\int \frac{d^3k}{(2\,\pi)^3}\,\biggl[\frac{\rho(\vec k^{\,2})}{\vec k^{\,2}}\,\biggl(\delta^{ij}-\frac{k^i\,k^j}{\vec k^{\,2}}\biggr)
+ \rho'(\vec k^{\,2})\,\frac{k^i\,k^j}{\vec k^{\,2}} \biggr] \,e^{i\,\vec k\cdot\vec r}\,. \label{50}
\end{align}
We thus obtain
\begin{align}
E_\mathrm{pole} =&\ \langle\phi|H_M |\phi\rangle\,, \label{51}
\end{align}
where $H_M$ is the nuclear recoil operator in the Breit approximation
\begin{align}
H_M = \frac{p^2}{2\,M} -\frac{1}{2\,M}\,\biggl\{p^j\,,\, \frac{Z\alpha}{2}\,\biggl[\frac{\delta^{ij}}{r} + \frac{r^i\,r^j}{r^3}\biggr]_\rho \alpha^i \biggr\}\,. \label{52}
\end{align}
It should be pointed out that
usage of $H_M$ with the finite-nuclear-charge distribution is not fully adequate,
since it leads to a numerically large spurious nuclear-size correction
at order $(Z\,\alpha)^5$, which will be examined in detail in the next section.

 \section{Recoil fns correction of order $\bm{(Z\,\alpha)^5}$}
 \label{sec:Za5recoil}
The $(Z\,\alpha)^5$ recoil fns correction consists of two parts. The first one is
the reduced-mass scaling of $\phi^2(0)$ in Eq.~(\ref{05}), and the second part is
given by the two-photon exchange amplitude. The sum is
\begin{align}
 E^{(5,1)}_{\rm fns} = &\ -3\,\frac{m}{M}\,E^{(5,0)}_\mathrm{fns} +
 \frac{i}{M}\,\phi^2(0)\,(4\,\pi\,Z\,\alpha)^2\! \int_s \frac{d^4k}{(2\,\pi)^4}
 \,G_{T}^{ij}(\omega,\vec{k})\,G_{T}^{ik}(\omega,\vec{k})
 \,\mathrm{Tr}\biggl[\gamma^j \,\frac{1}{\not\!t+\not\!k-m}\,\gamma^k\,\frac{(\gamma_0 + I)}{4}\biggr]
\nonumber \\ =&\ -3\,\frac{m}{M}\,E^{(5,0)}_\mathrm{fns} +
 \frac{i}{M}\,\phi^2(0)\,(4\,\pi\,Z\,\alpha)^2\! \int_s \frac{d^4k}{(2\,\pi)^4}\, [\rho^2(-k^2) -1]
\,\frac{\omega}{k^2+2\,m\,\omega} \,\biggl(\frac{1}{\omega^4} + \frac{2}{k^4}\biggr)\,. \label{53}
\end{align}
We perform the Wick rotation of the integration contour $\omega \rightarrow i\,\omega$ and
integrate first over the 3-dimensional sphere and then over $k$, obtaining
\begin{align}
  E^{(5,1)}_\mathrm{fns} = &\   -3\,\frac{m}{M}\,E^{(5,0)}_\mathrm{fns}
    -\frac{\phi^2(0)}{m\,M}\,(Z\,\alpha)^2
\nonumber \\ &\times
 \biggl[
\frac{3\,(1 - y^2)\,(1 + 4\,y^2 - 35\,y^4)}{8\,y^4}
- \frac{3 + 8\,y^2 + 40\,y^4 - 140\,y^6 + 105\,y^8}{16\,y^5}\,\ln\frac{1 + y}{1 - y} - \ln\frac{1 - y^2}{4}
 \biggr]
\,,
\end{align}
where $y = \sqrt{1-4\,m^2/\Lambda^2}$. The expansion of the above formula in small $m\,r_C$
reproduces the result obtained in Ref.~\cite{pachucki:23:prl},
\begin{align}
  E^{(5,1)}_\mathrm{fns} =
   -\frac{m}{M}\,\phi^2(0)\,(Z\,\alpha)^2\,\bigg( -\frac{43}{12} + \ln 12 - 2\,\ln m\,r_C \bigg)\,r_C^2
  -3\,\frac{m}{M}\,E^{(5,0)}_\mathrm{fns} + \frac{m^2}{M}\,O\big(m\,r_C\big)^4\,. \label{54}
\end{align}

We note that $E^{(5,1)}_\mathrm{fns}$ for electronic atoms depends on the nuclear radius as $r_C^2$.
If we were using the Breit approximation to derive the recoil fns correction of this order,
we would obtain a very different result, which is linear in the nuclear radius $r_C$
and numerically dominating. This fact was first pointed out by Friar in Ref.~\cite{friar:79:ap}.
The disappearance of the spurious $r_C$ term in a full-QED calculation was demonstrated
by Shabaev {\em et al.} \cite{shabaev:98:recground}.
Indeed, calculating the $\omega=0$ pole contribution in Eq. (\ref{53}), 
which corresponds to the Breit approximation, we obtain
\begin{align}
 E^{(5,1)}_{\rm rec,pole} = &\   -3\,\frac{m}{M}\,E^{(5,0)}_\mathrm{fns} +
 \frac{i}{M}\,\phi^2(0)\,(4\,\pi\,Z\,\alpha)^2\! \int \frac{d^4k}{(2\,\pi)^4}\,
\,\frac{(\rho^2(-k^2) -1)}{k^2+2\,m\,\omega} \, \frac{1}{2}\,\biggl(
\frac{1}{(\omega+\epsilon)^3} - \frac{1}{(\omega-\epsilon)^3}\biggr)
 \nonumber \\ =&\ -3\,\frac{m}{M}\,E^{(5,0)}_\mathrm{fns}
-\frac{m^2}{M}\,\frac{(Z\,\alpha)^5}{n^3}\,\frac{35}{32\,\sqrt{3}}\bigl(m\,r_C - 2\,m^3\,r_C^3\bigr) \,. \label{56}
\end{align}
The $E^{(5,1)}_{\rm rec,pole}$ contains
a spurious term $\sim r_C$, which is much larger than the correct one $\sim r_C^2\,\ln m\,r_C$.
The spurious term is only an artifact of an approximate treatment of the nuclear-recoil correction;
it disappears in the correct QED treatment.

We conclude that if one uses the relativistic recoil operator $H_M$ given by Eq.~(\ref{52})
(as it is routinely done in many-body calculations of atomic systems, see, e.g., Ref.~\cite{sahoo:24}),
one obtains unphysical results for the recoil fns correction.
In principle,
the spurious term can be removed by an additional correction to $H_M$. However, the coefficient
of this additional term depends on the exact forms of the Breit interaction for the extended size nucleus
and of the projector to the positive-energy subspace (used in the RMBPT or MCDF approaches).
So the additional correction to $H_M$ has to be carefully adjusted to the particular approximate calculations,
in order to remove this spurious linear in $r_C$ term. A better way is to account for the
QED recoil effects, e.g., by means of the model recoil operator \cite{anisimova:22}.

\section{ Recoil fns correction of order $\bm{(Z\,\alpha)^6}$}
We now derive the fns contribution of order $(\Za)^6$ coming from the recoil correction $E_{\rm rec}$
given by Eq.~(\ref{37}).
Let us split the $\omega$ integration contour in Eq. (\ref{37})
into the pole-contribution part $C_P$ and the high-energy part $C_H$.
$C_P$ encircles the  pole at $\omega=0$ and $C_H$
is the Wick-rotated integration contour along the negative real axis, which goes around the $\omega=0$ pole from the left,
see Fig.~\ref{fig:1}.
Accordingly, we split $E^{(6,1)}_\mathrm{fns}$ into three parts, assuming dimensional regularization
\begin{align}
E^{(6,1)}_\mathrm{fns}  = E^{(1)}_H + E^{(1)}_P + E^{(1)}_L\,. \label{57}
\end{align}
Here $E^{(1)}_H$ is the hard three-photon exchange with the $\omega$ integration carried out along the $C_H$ contour.
After subtracting the point-nucleus limit, the limit $\varepsilon \rightarrow 0$ can safely be approached.
The $E^{(1)}_P$ is the high-energy pole contribution, which is similar to the nonrecoil $E_H$ in Eq. (\ref{17}),
while $E^{(1)}_L$ is the low-energy pole contribution, which is similar to the nonrecoil $E_L$ in Eqs. (\ref{32}) and (\ref{33}).

The fns recoil high-energy part  $E^{(1)}_{H} $ is obtained from Eq. (\ref{37}) using the high-momentum three-photon exchange amplitude,
which yields
\begin{align}
E^{(1)}_{H} =&\ \frac{i}{M}\,(-4\pi Z\alpha)^3\,\phi^2(0) \int_{C_H} \frac{d\omega}{2\,\pi}\,
 \int\frac{d^dk_1}{(2\,\pi)^d}\,\int\frac{d^dk_2}{(2\,\pi)^d}
 \frac{\rho(-k_1^2)\,\rho(-k_2^2)\,\rho(\vec k_3^{\,2})-1}{k_1^2\,k_2^2\,\vec k_3^{\,2}}
 \nonumber \\ &\times
\biggl(\delta^{ik}-\frac{k_2^i\,k_2^k}{\omega^2}\biggr)\,\biggl(\delta^{jk}-\frac{k_1^j\,k_1^k}{\omega^2}\biggr)\,
 \mathrm{Tr}\biggl[\biggl(
 \gamma^i\,\frac{1}{\not\!k_2+\not\!t-m}\,\gamma^0\,\frac{1}{\not\!k_1+\not\!t-m}\,\gamma^j
 \nonumber \\ &\
 + \gamma^0\,\frac{1}{\not\!k_1 -\not\!k_2+\not\!t-m}\,\gamma^i\,\frac{1}{\not\!k_1+\not\!t-m}\,\gamma^j
 +  \gamma^i\,\frac{1}{\not\!k_2+\not\!t-m}\,\gamma^j\,\frac{1}{\not\!k_2 - \not\! k_1+\not\!t-m}\,\gamma^0
 \biggr)\,\frac{\gamma^0+I}{4}\biggr]\,, \label{58}
\end{align}
where $\vec k_3 = \vec k_1-\vec k_2$ and $k_i^2 = \omega^2-\vec k_i^2$.
We perform first $\int d^d k_1\,d^dk_2$ integration analytically using formulas from Appendix \ref{APPA},
and next integration over $\omega$ along the $C_H$ contour with the $\varepsilon\rightarrow 0$ limit. The result is
\begin{align}
E^{(1)}_{H} =&\  -\frac{m^2}{M}\,\frac{(Z\,\alpha)^6}{n^3}\, f(m\,r_C)\,,
\label{59b}
\end{align}
where
\begin{align}
f(m\,r_C)=&\
\frac{m\,r_C}{2\,\sqrt{3}}\,\biggl( \frac{13}{6} - \frac{64}{9\,\sqrt{3}} - \frac{350}{9\,\pi} + \frac{80\,\ln 2}{\pi}\biggr)
+\frac{m^2\,r_C^2}{12}\,\biggl( \frac{1895}{216} + 64\,\ln 2  \biggr) + (m\,r_C)^3\,\delta f(m\,r_C)\,. \label{59}
\end{align}
The function $\delta f(m\,r_C)$ was computed numerically with the help of Wolfram {\sl Mathematica}. The contribution
of $\delta f$ is negligible for electronic atoms but significant for the muonic atoms.
Table~\ref{table_f} lists numerical values of $\delta f(m\,r_C)$ for the range of argument
$0.45 \leq m_\mu\,r_C \leq 2.40$ relevant for muonic atoms.

\begin{table}
\caption{Numerical values of the function $\delta f(x)$ defined by Eq.~(\ref{59}).}
\label{table_f}
\begin{ruledtabular}
\begin{tabular}{w{3.2}w{3.5}w{3.2}w{3.5}w{3.2}w{3.5}w{3.2}w{3.5}w{3.2}w{3.5}w{3.2}w{3.5}}
\multicolumn{1}{c}{$x$} & \multicolumn{1}{c}{$\delta f(x)$} &
\multicolumn{1}{c}{$x$} & \multicolumn{1}{c}{$\delta f(x)$} &
\multicolumn{1}{c}{$x$} & \multicolumn{1}{c}{$\delta f(x)$} &
\multicolumn{1}{c}{$x$} & \multicolumn{1}{c}{$\delta f(x)$} &
\multicolumn{1}{c}{$x$} & \multicolumn{1}{c}{$\delta f(x)$}
\\
\hline\\[-5pt]
0.45 &    -3.378\,26 & 0.85 &   -2.41382 & 1.25   &  -1.90976  & 1.65   &  -1.59053   &  2.05  &  -1.36749\\
0.50 &    -3.208\,65 & 0.90 &   -2.33488 & 1.30   &  1.86238   &  1.70  &  -1.55846   &  2.10  &  -1.34418\\
0.55 &   -3.058\,29 & 0.95  &   -2.26153 & 1.35   &  -1.81749  &  1.75  &  -1.52775   &  2.15  &  -1.32169\\
0.60 &   -2.923\,74 & 1.00  &   -2.19315 & 1.40   &  -1.77489  &  1.80  &  -1.49831   &  2.20  &  -1.29998\\
0.65 &   -2.802\,38 & 1.05  &   -2.12921 & 1.45   &  -1.73440  &  1.85  &  -1.47005   &  2.25  &  -1.27901\\
0.70 &   -2.692\,16 & 1.10  &   -2.06926 & 1.50   &  -1.69585  &  1.90  &  -1.44290   &  2.30  & -1.25874\\
0.75 &   -2.591\,49 & 1.15  &   -2.01292 & 1.55   &  -1.65911  &  1.95  &  -1.41680   &  2.35  &  -1.23914\\
0.80 &   -2.499\,06 & 1.20  &   -1.95985 & 1.60   &  -1.62404  &  2.00  &  -1.39168   &  2.40  &  -1.22017\\
\end{tabular}
\end{ruledtabular}
\end{table}

Before considering the fns recoil pole high-energy part $E^{(1)}_P$,
let us calculate the reduced-mass correction to $\phi^2(0)$ in the dimensional regularization.
The nonrelativistic Hamiltonian in atomic units is
\begin{equation}
H_\mu = \frac{\vec{p}^{\,2}}{2} + \frac{\vec{p}^{\,2}}{2}\,\frac{m}{M}
-\frac{C_1}{r^{1-2\,\epsilon}}\,. \label{60}
\end{equation}
Variable change $ \vec r \rightarrow \vec r\,\mu^{-1/(1+2\,\epsilon)}$
leads to $H_\mu \rightarrow \mu^\frac{1-2\,\epsilon}{1+2\,\epsilon}\,H  \approx (1- \frac{m}{M}\,(1-4\,\epsilon))\,H$,  thus
\begin{align}
\phi^2(0) = \langle\phi| \delta^d(r) |\phi\rangle \rightarrow \mu^\frac{3-2\,\epsilon}{1+2\,\epsilon}\,\phi^2(0)
\approx \mu^{3-8\,\epsilon}\,\phi^2(0)
\approx \biggl(1- (3-8\,\epsilon)\,\frac{m}{M}\biggr)\,\phi^2(0)\,.  \label{61}
\end{align}

We now examine the pole contribution, which  is split into two parts:
\begin{align}
E^{(1)}_P =&\ E_{P1} + E_{P2}\,.  \label{62}
\end{align}
The first part $E_{P1}$ is induced by $p^2/(2\,M)$ and evaluated as
\begin{align}
 E_{P1} =&\ -(3-8\,\epsilon)\,\frac{m}{M}\,E_H -(4\,\pi\,Z\,\alpha)^3\,\phi^2(0)\,\int \frac{d^dq_1}{(2\,\pi)^d}\,\int \frac{d^dq_2}{(2\,\pi)^d}\,
  \frac{\rho(q_1^2)}{q_1^2}\, \frac{\rho(q_2^2)}{q_2^2}\, \frac{\rho(q_3^2)}{q_3^2}\,  \frac{1}{2\,M}
  \nonumber \\ &\times
 \biggl\{ \vec q_1^{\,2}\,{\rm Tr}\biggl[\frac{1}{(\not\!p_1-m)}\,\gamma_0\, \frac{1}{(\not\!p_1-m)}\,
  \gamma_0\, \frac{1}{(\not\!p_2-m)}\, \frac{(\gamma_0+I)}{4}\biggr] + (1\leftrightarrow 2)\biggr\}
  \nonumber \\ =&\ -(5-8\,\epsilon)\,\frac{m}{M}\,E_{H1}  -(3-8\,\epsilon)\,\frac{m}{M}\,E_{H2}
  \nonumber \\ &\
  +(4\,\pi\,Z\,\alpha)^3\,\phi^2(0)\,\int \frac{d^dq_1}{(2\,\pi)^d}\,\int \frac{d^dq_2}{(2\,\pi)^d}\,
  \frac{\rho(q_1^2)}{q_1^4}\, \frac{\rho(q_2^2)}{q_2^4}\, \frac{\rho(q_3^2)}{q_3^2}\,  \frac{m}{M}\,
 (\vec q_1+\vec q_2)^2 \,, \label{63}
 \end{align}
 where $E_H$, $E_{H1}$, and $E_{H2}$ are defined in Eqs. (\ref{18}), (\ref{19}), and (\ref{20}), respectively.
The second term $E_{P2}$ is due to the electron-nucleus Breit interaction. It is expressed as
\begin{align}
E_{P2} =&\ (4\,\pi\,Z\,\alpha)^3\,\phi^2(0)\,\int \frac{d^dq_1}{(2\,\pi)^d}\,\int \frac{d^dq_2}{(2\,\pi)^d}\,
 \frac{\rho(q_1^2)}{q_1^2}\, \frac{\rho(q_2^2)}{q_2^2}\, \frac{\rho(q_3^2)}{q_3^2}\,\frac{1}{2\,M}
  \nonumber \\ &\times \biggl\{
{\rm Tr}\biggl[\gamma^i\,\frac{1}{(\not\!p_1-m)}\,
  \gamma_0\, \frac{1}{(\not\!p_2-m)}\,\gamma_0\, \frac{(\gamma_0+I)}{4}\biggr]\,
  q_1^j\,G_{C}^{ij}(0,-\vec q_1)\, \frac{q_1^2}{\rho(q_1^2)}
  \nonumber \\ &\ +
  {\rm Tr}\biggl[\gamma_0\,\frac{1}{(\not\!p_1-m)}\,
  \gamma^i\, \frac{1}{(\not\!p_2-m)}\,\gamma_0\, \frac{(\gamma_0+I)}{4}\biggr]\,
(q_1^j + q_2^j)\,G_{C}^{ij}(0,\vec q_3)\,\frac{q_3^2}{\rho(q_3^2)}
  \nonumber \\ &\ +
  {\rm Tr}\biggl[\gamma_0\,\frac{1}{(\not\!p_1-m)}\,
  \gamma_0\, \frac{1}{(\not\!p_2-m)}\,\gamma^i\, \frac{(\gamma_0+I)}{4}\biggr]\,
q_2^j\,G_{C}^{ij}(0,\vec q_2)\, \frac{q_2^2}{\rho(q_2^2)}\biggr\}\,.  \label{64}
\end{align}
where $\vec q_3 = \vec q_1-\vec q_2$.
After performing traces we obtain
\begin{align}
E_{P2} =&\
-(4\,\pi\,Z\,\alpha)^3\,\phi^2(0)\,\int \frac{d^dq_1}{(2\,\pi)^d}\,\int \frac{d^dq_2}{(2\,\pi)^d}\,
\frac{\rho(q_1^2)}{q_1^4}\, \frac{\rho(q_2^2)}{q_2^4}\, \frac{\rho(q_3^2)}{q_3^4}\,\frac{1}{M}
  \nonumber \\ &\times
\biggl\{
4\,q_1^2\,q_2^2 -4\,(\vec q_1\cdot\vec q_2)^2
+ \frac{\rho'(q_3^{2})}{\rho(q_3^2)}\,
q_3^2\,\bigl[(q_1^2-q_2^2)^2\ + q_2^2\,q_3^2 + q_1^2\,q_3^2\bigr]  \label{65}
\biggr\}\,.
\end{align}
The  complete fns recoil pole high-energy part $E^{(1)}_P = E_{P1} + E_{P2}$  is
\begin{align}
E^{(1)}_P =&\
-(5-8\,\epsilon)\,\frac{m}{M}\,E_{H1}  -(3-8\,\epsilon)\,\frac{m}{M}\,E_{H2}  + \delta E_P \,, \label{66}
\end{align}
where
\begin{align}
\delta E_P =&\
-(4\,\pi\,Z\,\alpha)^3\,\phi^2(0)\,\frac{m}{M}\,\int \frac{d^dq_1}{(2\,\pi)^d}\,\int \frac{d^dq_2}{(2\,\pi)^d}\,\frac{\partial}{\partial q_3^2}
\frac{\rho(q_1^2)}{q_1^4}\, \frac{\rho(q_2^2)}{q_2^4}\, \rho(q_3^{2})\,\biggl[ \frac{(q_1^2-q_2^2)^2}{q_3^2} + q_2^2 + q_1^2\biggr]
\nonumber \\ =&\
\frac{(Z\,\alpha)^6}{n^3}\,\frac{m^2}{M}\,\frac{2471}{2592}\,m^2\,r_C^2\,.  \label{67}
\end{align}

The low-energy part $E^{(1)}_L$, with $|\vec k|\sim m\,\alpha$, is obtained by using  the nonrelativistic expansion of $H_M$ in Eq. (\ref{52}).
Using
\begin{align}
-\frac{1}{8\,\pi}\,\biggl[\frac{\delta^{ij}}{r} + \frac{r^i\,r^j}{r^3}\biggr]_\rho =&\
-\frac{1}{8\,\pi}\,\biggl[\frac{\delta^{ij}}{r} + \frac{r^i\,r^j}{r^3}\biggr]_\epsilon -\rho'(0)\,\delta^d(r)\,,  \label{68}
\end{align}
we obtain
\begin{align}
H_M =&\ \frac{m}{M}\,\biggl[
\frac{p^2}{2} - \frac{1}{2}\,p^i\,\bigg(\frac{\delta^{ij}}{r} + \frac{r^i r^j}{r^3}\bigg)_\epsilon\,p^j  + \frac{1}{4}\,\big\{ p^i , \big\{ p^i , \delta V\big\}\!\big\}\biggr]\,.  \label{69}
\end{align}
This expression should be combined with the nonrelativistic expansion of the Dirac-Coulomb Hamiltonian in Eq. (\ref{29}) to obtain
\begin{align}
H_{D} +H_M =&\ H_\mu -\frac{p^4}{8} + \frac{\pi}{2}\,\delta^d(r)  + \delta V + \delta^{(2)}V + \frac{1}{8}\,\nabla^2(\delta V) +
\frac{m}{M}\,\biggl[ - \frac{1}{2}\,p^i\,\bigg(\frac{\delta^{ij}}{r} + \frac{r^i r^j}{r^3}\bigg)_\epsilon\,p^j  + \frac{1}{4}\,\big\{ p^i , \big\{ p^i , \delta V\big\}\!\big\}\biggr]\,. \label{71}
\end{align}
After rescaling the reduced mass, this expression becomes
\begin{align}
H_{D} +H_M =&\
\biggl(1-\frac{m}{M}\,(1-4\,\epsilon)\bigg)\,H -\biggl(1-\frac{m}{M}\,(4-8\,\epsilon)\bigg)\,\frac{p^4}{8}
+ \biggl(1-\frac{m}{M}\,(3-8\,\epsilon)\bigg)\,\biggl( \frac{\pi}{2}\,\delta^d(r)  + \delta V\biggr)
\nonumber \\ &\
+ \bigg(1-\frac{m}{M}\,5\bigg)\,\bigg(\delta^{(2)}V + \frac{1}{8}\,\nabla^2(\delta V)\bigg)
+ \frac{m}{M}\,\biggl[ - \frac{1}{2}\,p^i\,\bigg(\frac{\delta^{ij}}{r} + \frac{r^i r^j}{r^3}\bigg)_\epsilon\,p^j  + \frac{1}{4}\,\big\{ p^i , \big\{ p^i , \delta V\big\}\!\big\}\biggr]\,. \label{72}
\end{align}
Similarly to the derivation of the nonrecoil fns correction,  the low-energy part ${\cal E}^{(1)}_L$ is split into two parts,
\begin{align}
  {\cal E}^{(1)}_L =&\ {\cal E}^{(1)}_{L1} + {\cal E}^{(1)}_{L2}\,, \label{73}
\end{align}
where ${\cal E}^{(1)}_{L1}$ is the nonrelativistic contribution proportional to $r_C^4$,
\begin{align}
   {\cal E}^{(1)}_{L1} =&\ \langle\delta V\,\frac{1}{(E-H)'}\,\delta V\rangle  + \langle\delta^{(2)} V\rangle
 =   -(5-12\,\epsilon)\,\frac{m}{M}\,{\cal E}_{L1}\,, \label{74}
 \end{align}
  and ${\cal E}^{(1)}_{L2}$ is the relativistic part proportional to $r_C^2$
\begin{align}
   {\cal E}^{(1)}_{L2} =&\
\frac{m}{M}\,\bigg\{
-\frac{5}{8}\,\langle \nabla^2(\delta V)  \rangle
    + 2\,(6-12\,\epsilon)\,\bigg\langle \delta V\frac{1}{(E-H)'} \frac{p^4}{8}  \bigg\rangle
     - 2\, (5-12\,\epsilon) \bigg\langle \delta V\frac{1}{(E-H)'}  \frac{\pi}{2}\,\delta^d(r) \bigg\rangle
   \nonumber \\ &\
   +   \frac{1}{4}\,\langle \big\{ p^i , \big\{ p^i , \delta V\big\}\!\big\} \rangle
  - 2\, \bigg \langle \delta V\frac{1}{(E-H)'} \frac{1}{2}\,p^i\,\bigg[\frac{\delta^{ij}}{r} + \frac{r^i r^j}{r^3}\bigg]_\epsilon\,p^j  \bigg \rangle
\biggr\}
\nonumber \\ =&\
-(3-12\,\epsilon)\,\frac{m}{M}\,{\cal E}_{L2}  + \frac{m^3}{M}\,r_C^2\,\frac{1}{n^3}\,\biggl(\frac{1}{n^2} +\frac23 \biggr)
\,. \label{75}
\end{align}
We observe here an additional state dependence, beyond that in ${\cal E}_{L2}$.
The sum of all calculated recoil terms is
\begin{align}
E^{(6,1)}_\mathrm{fns} =&\ E^{(1)}_H + E^{(1)}_P + E^{(1)}_L
=
-\frac{m^2}{M}\,\frac{\partial}{\partial m} E^{(6,0)}_\mathrm{fns}
+ \frac{m^2}{M}\,\frac{(Z\,\alpha)^6}{n^3}\,\biggl[m^2\,r_C^2\,\biggl(\frac{1}{n^2} +\frac{4199}{2592} \biggr) - f(m\,r_C)\biggr]\,, \label{76}
\end{align}
where the function $f$ is defined in Eq.~(\ref{59b}).
In the above, the first term corresponds to the reduced-mass rescaling of the corresponding nonrecoil correction
and the second one is a remainder. We note that the function $f$ contains a contribution linear in $r_C$
[see Eq.~(\ref{59})], which is numerically large
for electronic atoms. As a result, the recoil fns contribution of order $(\Za)^6$ 
numerically dominates over the $(\Za)^5$ contribution
for electronic atoms, a surprising fact, first found in Ref.~\cite{pachucki:23:prl} based on numerical calculations.

\section{Total fns recoil correction}

We now summarize all known contributions to the fns effect of light hydrogenlike atoms for a finite-mass nucleus,
$E_{\rm fns} = E^{(4)}_{\rm fns} + E^{(5)}_{\rm fns} + E^{(6)}_{\rm fns}$.
The leading-order nuclear contribution is
\begin{equation}
  E^{(4)}_{\rm fns} =  \frac{2\,\pi}{3}\,\frac{(\Za)^4}{\pi n^3}\,\mu^3\,r_C^2\,
  \delta_{l,0}\,,
  \label{e72}
\end{equation}
where $\mu = mM/(m+M)$ is the reduced mass.
The above formula is model independent and includes the nuclear recoil effect to all orders in $m/M$.

The finite-nuclear-size contribution of order $(\Za)^5$, calculated to first order in $m/M$ but to all orders
in $mr_C$, is given by
\begin{align}
  E^{(5)}_\mathrm{fns} = &\
  -m\,\frac{(Z\,\alpha)^5}{\pi n^3}\,\mu^3\,\delta_{l,0}
  \biggl\{
  \frac{\pi\,r^3_F}{3}
+ \frac{1}{M\,m^2} \biggl[
\frac{3\,(1 - y^2)\,(1 + 4\,y^2 - 35\,y^4)}{8\,y^4}
  \nonumber \\ &
- \frac{3 + 8\,y^2 + 40\,y^4 - 140\,y^6 + 105\,y^8}{16\,y^5}\,\ln\frac{1 + y}{1 - y} - \ln\frac{1 - y^2}{4}
 \biggr] \biggr\}\,,
\end{align}
where $y = \sqrt{1-(mr_C)^2/3}$. This contribution is model dependent. 
The model dependence of the nonrecoil part can be estimated by comparing
results obtained in the exponential and Gaussian models with
the corresponding results for $r_F$ summarized in Table~\ref{tab:models}.
The recoil part of $E^{(5)}_\mathrm{fns}$ is calculated within the exponential nuclear model only. 
In the case of an arbitrary nuclear model, the above formulas are not valid and 
one has to return to Eq. (\ref{53}). For the integral over $k$ in Eq. (\ref{53}) to exist, the
nuclear form factor $\rho(-k^2)$ should not only be an analytical function in the complex plane
(with a branch cut on the negative axis), but also  vanish at complex infinity.
We note that many popular nuclear models (e.g., Gaussian and Fermi) do not satisfy this condition
and are thus not suitable for describing the recoil fns effect.
 
For the electronic atoms, Eq. (73) can be simplified by using the fact that the parameter
$mr_C$ is small. Expanding in $mr_C$, we obtain
\begin{align}
E^{(5)}_\mathrm{fns} =&\ -m\,\frac{(Z\,\alpha)^5}{n^3}\,\mu^3\,\,\delta_{l,0}
\biggl[\frac{r^3_F}{3}
-\frac{1}{\pi}\,\bigg( \frac{43}{12} - \ln 12 + 2\,\ln m\,r_C \bigg)\,\frac{r_C^2}{M}  + \frac{1}{m^2 M}\,O\big(m\, r_C\big)^4 \biggr]\,.
\end{align}
The dependence on $r_C$ is the same for all the models and only the coefficient $43/12 -\ln12$ is a subject of the model dependence.

The finite size effect at order $(\Za)^6$, calculated
for $nS$ states to first order in $m/M$ but to all orders in $mr_C$, is given by
\begin{align}
E^{(6)}_\mathrm{fns}(nS) =&
-(Z\,\alpha)^6\,\mu^3\,r_C^2\,\frac{2}{3\,n^3}\,\biggl[-\frac{5}{4} + \frac{9}{4\,n^2} -\frac{1}{n}  -\ln n+\gamma+\Psi(n)
        +\kappa_1 +\ln(\mu\,r_{C}\,Z\,\alpha)\biggr]
        \nonumber \\
        & + (Z\,\alpha)^6\,\mu^5\,r_C^4\,\frac{4}{9\,n^3}\,\biggl[1-\frac{1}{n}  - \ln n  + \gamma + \Psi(n) +\kappa_2
        + \ln(\mu\,r_C\,Z\,\alpha)\biggr]
        \nonumber \\ &
        +(Z\,\alpha)^6\,\mu^5\,r_{CC}^4\,\frac{1}{15\,n^5}
        +\frac{m^2}{M}\,\frac{(Z\,\alpha)^6}{n^3}\,\biggl[m^2\,r_C^2\,\biggl(\frac{1}{n^2} + \frac{4199}{2592} \biggr) -f(m\,r_C)\biggr]\,, \label{77}
\end{align}
where $r_{CC}$ is defined by Eq.~(\ref{13a}),
$\kappa_1$ and $\kappa_2$ are defined in Eqs. (\ref{35}) and (\ref{36}), respectively, and $f(m\,r_C)$ is defined in Eq. (\ref{59}),
with numerical values summarized in Table~\ref{table_f}.
The contribution $E^{(6)}_\mathrm{fns}$ is model dependent, with the model dependence coming through
$r_{CC}$, $\kappa_1$, $\kappa_2$, and $f(mr_C)$.
The model dependence of the nonrecoil part can be estimated by comparing
results obtained in the exponential and Gaussian models, with the help of
results summarized in Table~\ref{tab:models}. The recoil part of
$E^{(6)}_\mathrm{fns}$ is calculated for the exponential nuclear model only.
For electronic atoms, $E^{(6)}_\mathrm{fns}$ can be expanded in $mr_C$, with the result
 \begin{align}
E^{(6)}_\mathrm{fns}(nS) = &
-(Z\,\alpha)^6\,\mu^3\,r_C^2\,\frac{2}{3\,n^3}\,\biggl[-\frac{5}{4} + \frac{9}{4\,n^2} -\frac{1}{n}  -\ln n+\gamma+\Psi(n)
        +\kappa_1 +\ln(\mu\,r_{C}\,Z\,\alpha)\biggr]
        \nonumber \\ &
        +\frac{m^2}{M}\,\frac{(Z\,\alpha)^6}{n^3}\,\biggl[
 -\frac{m\,r_C}{2\,\sqrt{3}}\,\biggl( \frac{13}{6} - \frac{64}{9\,\sqrt{3}} - \frac{350}{9\,\pi} + \frac{80\,\ln 2}{\pi}\biggr)
 + m^2\,r_C^2\,\biggl(\frac{1}{n^2} + \frac{8}{9}  - \frac{16}{3}\,\ln 2  \biggr) + O\big(m\,r_C\big)^3 \biggr]
 \,, \label{78}
\end{align}
where the second line includes all terms beyond the reduced-mass scaling of the nonrecoil relativistic correction.
Regarding model dependence we expect the same functional form, while
the constant $\kappa_1$ and value of coefficients of  $r_C$ and $r^2_C$ in the second line of Eq. (\ref{78}) are model dependent.

We would like to point out that for electronic atoms, the $(\Za)^5$ fns correction is very small numerically
(since it is suppressed by both $\Za$ and a small parameter $mr_C$ as compared to the leading fns effect).
On the contrary, the $(\Za)^6$ fns
contribution, being enhanced by $\ln(mr_C\Za)$, is numerically significant and gives the dominant
correction to the leading fns correction (\ref{e72}).

For the $P$ states, the finite size corrections $E^{(4)}_{\rm fns}$  and $E^{(5)}_{\rm fns}$ vanish,
while at the  $(Z\,\alpha)^6$ order they take the form \cite{patkos:24}
\begin{align}
E^{(6)}_{\rm fns}(nP_J) =&\ (Z\,\alpha)^6\, \frac{1}{9}\,\biggl(\frac{1}{n^3}-\frac{1}{n^5}\biggr)
\biggl[ \biggl(\frac{1}{2} - \langle\vec L\cdot\vec s\,\rangle\biggr)\,\mu^3\,r^2_C
+ \frac{1}{5}\,\mu^5\,r^4_{CC}\biggr]\,, \label{08}
\end{align}
where $\langle \vec L\cdot\vec s \rangle = -1,\nicefrac{1}{2}$ for $J=\frac{1}{2}, \frac{3}{2}$, respectively.
This simple formula is valid up to the first order in  $m/M$ mass ratio;
the general result valid to all orders in mass ratio was obtained in Ref.~\cite{patkos:24}.
The $E^{(6)}_{\rm fns}$ vanishes for states with $L > 1$.

Formulas summarized in this section demonstrate that the dependence of the fns effect on the
nuclear charge radius $r_C$ is generally quite complicated. It contains odd powers
of $r_C$, logarithms, and model-dependent parameters.
This makes it abundantly clear that the use  of the so-called Seltzer moments \cite{seltzer:69}, which
assumes expansion of the fns correction to energies in even multipoles of $r_C$,
is incorrect.
It is truly astonishing that the expansion in terms of Seltzer moments, despite lacking any
physical justification, continues to appear in the literature to this day
\cite{hur:22,sahoo:24}.
\end{widetext}

\begin{table*}
  \caption{Results for the exponential and the Gaussian models of the nuclear charge distributions.
\label{tab:models}
}
\begin{ruledtabular}
\begin{tabular}{lcc}
\multicolumn{1}{l}{Quantity} & \multicolumn{1}{c}{Exponential} & \multicolumn{1}{c}{Gaussian}
\\
\hline
\\[-5pt]
$\rho(q^2)$   & $\Lambda^4/(\Lambda^2+q^2)^2$ & $\exp\bigl(\frac{a\,q^2}{2}\bigr)$\\[2pt]
$\rho(r)$     & $\Lambda^3/(8\,\pi)\,e^{-\Lambda\,r}$ &  $(2\pi a)^{-3/2}\, {\exp\bigl(- \frac{r^2}{2\,a}\bigr)}$ \\[2pt]
$V(r)$      & $(1/r) \bigl(1 - e^{-\Lambda\,r}\bigr) - (\Lambda/2)\,e^{-\Lambda\,r}$
              & $(1/r)\,{\rm erf}\bigl(\frac{r}{\sqrt{2\,a}}\bigr)$ \\[2pt]
$r_C$         & $2\,\sqrt{3}/\Lambda$ & $\sqrt{3\,a}$  \\[2pt]
  $r_{CC}/r_C$ & $1.257\,433$ & $1.136\,219$ \\
  $r_{F}/r_C$  & $1.558\,965$ & $1.514\,599$ \\
  $\kappa_1$  & \!\!$-0.413\,384$ & \!\!$-0.465\,457$ \\
  $\kappa_2$  & $2.356\,581$ & $1.688\,528$
 \end{tabular}
\end{ruledtabular}
\end{table*}

\section{Results and Discussion}

In this paper we extended our previous studies of the fns effect
for electronic and muonic hydrogenlike atoms \cite{pachucki:18} by calculating the nuclear recoil fns corrections
of orders $(\Za)^5$ and $(\Za)^6$. Calculations were performed without any expansion in the
nuclear-charge radius $r_C$, ensuring that the obtained results are applicable for muonic atoms.

\begin{figure}[t]
\centerline{\resizebox{0.5\textwidth}{!}{\includegraphics{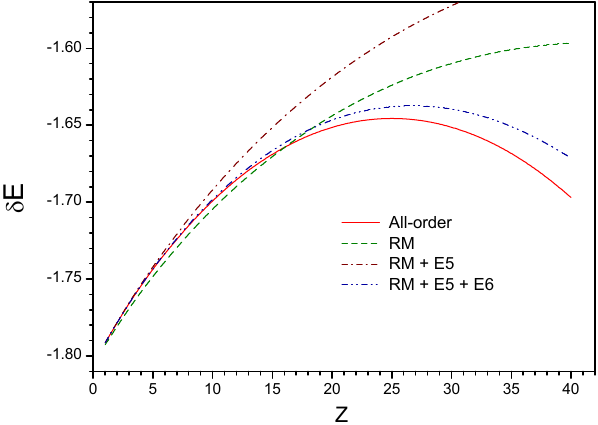}}}
 \caption{
 Finite-size recoil correction, calculated to all orders in $\Za$ (red solid line) and within the
 $\Za$ expansion, for the $1s$ state of muonic atoms with the nuclear charge radius fixed at $r_C = 1$~fm.
 Units are $\delta E/[(m^2/M) (\Za)^4/\pi]$.
 The green dashed line corresponds to the reduced-mass (RM) contribution
 $-3 (m/M) E_{\rm fns}$ coming from $\phi^2(0)$. The brown dash-dotted line shows results including
 the $(\Za)^5$ correction (E5), whereas the blue dash-double-dotted line shows results including the $(\Za)^6$ correction (E6).
   \label{fig:recfns}}
\end{figure}

The derived formulas were checked by comparing with numerical calculations performed to
all orders in $\Za$ by the method described in Ref.~\cite{yerokhin:23:recoil}. The comparison
of the all-order and the $\Za$-expansion results is  shown in Fig.~\ref{fig:recfns} for the $1S$ state of
muonic atoms. In order to obtain smooth curves, we artificially
fixed the nuclear-charge radius at $r_C = 1$~fm for all nuclear charges. We observe that
the $\Za$-expansion results converge to the all-order values in the limit of $Z\to 0$.
It is remarkable that the inclusion of the $(\Za)^6$ correction is essential in order
to achieve good agreement with all-order results for small values of $Z$.

Now we examine the magnitude of the obtained corrections for several cases that are interesting from the
experimental point of view.
The first case is the $2S$-$1S$ transition in hydrogen, which
is the basis of the determination of the Rydberg constant and
has been measured with 10~Hz accuracy \cite{parthey:11, matveev:13}.
We obtain the following results for the $(\Za)^5$ and $(\Za)^6$ fns corrections:
\begin{align}
E^{(5,0)}_\mathrm{fns}(\mathrm{H}, 2S\mathrm{-}1S) =&\  29\;\mathrm{Hz}\,, \nonumber \\
E^{(5,1)}_\mathrm{fns}(\mathrm{H}, 2S\mathrm{-}1S) =&\  20\;\mathrm{Hz}\,, \nonumber \\
E^{(6,0)}_\mathrm{fns}(\mathrm{H}, 2S\mathrm{-}1S) =&\  -585\,(5)\;\mathrm{Hz}\,, \nonumber \\
E^{(6,1)}_\mathrm{fns}(\mathrm{H}, 2S\mathrm{-}1S) =&\  20\;\mathrm{Hz}\,. \label{79}
\end{align}
We see that the $E^{(5,0)}_\mathrm{fns}$ correction is abnormally small, being
suppressed by an additional power of $r_C$.
It is interesting that the recoil $(\Za)^5$ correction has nearly the same magnitude as
the nonrecoil contribution to this order.
We note that for hydrogen the $(\Za)^5$ corrections are already included in the
nuclear-structure calculations, performed, for example,  in Ref. \cite{tomalak:19}.
The uncertainty of $E^{(6,0)}_\mathrm{fns}$ is the estimated model dependence, evaluated as
twice the difference between the exponential and the Gaussian models. This model-dependence
uncertainty is comparable to the theoretical uncertainty due to $\alpha^5$ nuclear
polarizability of 11~Hz \cite{tomalak:19}.
The corrections in Eq.~(\ref{79}) are larger than experimental uncertainty but smaller than the other
theoretical uncertainties (of about 1~kHz) and the proton-radius uncertainty (about 1~kHz)
\cite{mohr:24:codata}.

Next we analyze the $2P$-$2S$ transition in light muonic atoms. Our numerical
results for the recoil fns correction for
muonic atoms with $Z = 1$ and $Z = 2$ are presented in Table \ref{table_mu}.
\begin{table}[t]
\renewcommand{\arraystretch}{0.92}
\caption{$E^{(6,1)}_\mathrm{fns}$ for the $2P$-$2S$ transition in
light muonic atoms, in $\mu$eV.}
\label{table_mu}
\begin{ruledtabular}
\begin{tabular}{llw{2.3}w{2.3}w{2.3}}
Atom&\multicolumn{1}{c}{$r_C$[fm]} &   \multicolumn{1}{c}{$m_\mu\,r_C$}
&  \multicolumn{1}{c}{$f(m_\mu\,r_C)$} &  \multicolumn{1}{c}{$E^{(6,1)}_\mathrm{fns}(2P-2S)$}\\[0.5ex] \hline
$\mu$H & $0.840\,60(39)$       &   0.450	&	   1.022 & 0.69	\\
$\mu$D & $2.127\,58(78)$       &    1.139   &        3.849  & 1.18   \\	
$\mu^3$He & $1.970\,07(94)$ &    1.055   &        3.450   & 48.7   \\
$\mu^4$He & $1.678\,6(12)$   &   0.899   & 	  2.745   &  29.2 \\
\end{tabular}
\end{ruledtabular}
\end{table}
Recoil fns results for $\mu$H and $\mu$D are smaller than experimental uncertainty
(of $2.3$ and $3.4\;\mu$eV, respectively) and thus negligible at present.
For the helium isotopes, they are close to the experimental uncertainty
($48\;\mu$eV) but still smaller than the uncertainty from
the unknown nuclear-polarizability effects in the three-photon exchange.
One may expect that the recoil fns corrections
become more significant for heavier elements, such as $\mu$Li or $\mu$Be,
due to enhancement by $Z^2$ with respect to
the leading finite size.

It should be stressed that the fns corrections are only the elastic part of the total
nuclear-structure effect. The remaining, inelastic part includes
the nuclear polarizability effect. In principle,
it is advantageous to account for the elastic and
inelastic nuclear-structure parts on the same footing, but this is not always possible.
The nuclear-structure effects
have been accurately calculated at the order $(Z\,\alpha)^5$ for H and He isotopes,
for both electronic and muonic atoms (see Ref.  \cite{pachucki:24:rmp} and references therein), 
and estimated for heavier muonic atoms \cite{gorchtein:25}.
However, the inelastic $(Z\,\alpha)^6$ correction is not known yet and is the source of the
main theoretical uncertainty in light muonic atoms.

\section{Summary}

We derived formulas for the nuclear-recoil fns corrections
of orders $(\Za)^5$ and $(\Za)^6$. Our calculations were performed without any expansion in the
nuclear-charge radius $r_C$, which makes the obtained results applicable for both electronic and
muonic atoms. The obtained results are relevant for high-precision determinations of the
root-mean-square charge radii from spectroscopy of muonic atoms \cite{ohayon:24}.

We demonstrated that the application of the widely-used Breit approximation to the nuclear recoil effect
with an extended nuclear charge distribution
leads to the appearance of an unphysical fns contribution
at the $(Z\,\alpha)^5$ order, which is linear in the nuclear-charge radius $r_C$.
This spurious term disappears in the full-QED treatment, leaving 
the correct contribution $\propto r_C^2\,\ln r_C$, which is, for the electronic atoms,
much smaller than this spurious term.

The recoil fns correction contributes to the nonlinearity of the so-called King
plots in the isotope-shift measurements of many-electron atoms \cite{hur:22,wilzewski:24},
which are used for searches of new-physics scalar boson fields coupling to electrons
and neutrons. This effect should be included in the theoretical analysis of the observed nonlinearities.
It is interesting that the leading recoil fns effect $\sim (\Za)^4$ does not contribute to the King plot nonlinearities
(since the reduced-mass prefactor can be effectively absorbed into the nuclear radius). 
Therefore, it is essential that the full QED approach
is used for description of the recoil fns effect when studying these nonlinearities.

A possible future application of the method developed in this work would be
a calculation of the analogous correction to the hyperfine splitting in $\mu$H,
which could provide a sensitive low-energy test of the Standard Model \cite{nuber:23,vacchi:23}
by comparing the splitting intervals in electronic and muonic atoms.
Further application could be a calculation of the radiative recoil fns correction, which can be
as large as that obtained in this work.
Yet another application could be a calculation of $(Z\,\alpha)^7$ and $\alpha(Z\,\alpha)^6$
recoil corrections to the Lamb shift and hyperfine splitting in hydrogenic systems, which
are not known and limit the accuracy of the current theoretical predictions \cite{mohr:24:codata}.

\appendix

\section{$\bm{d^dk}$ integration}
\label{APPA}
We aim to perform the following integration in $d=3-2\,\epsilon$  dimensions, assuming  $\epsilon$ to be small:
\begin{align}
f(m_1,m_2,m_3; n_1, n_2, n_3) = \int\frac{d^d k_1}{(2\,\pi)^d}\, \int\frac{d^d k_2}{(2\,\pi)^d}
\nonumber \\
\times\frac{4\,\pi}{(k_1^2 + m_1^2)^{n_1}}\, \frac{4\,\pi}{(k_2^2 + m_2^2)^{n_2}}\,\frac{4\,\pi}{(k_3^2 + m_3^2)^{n_3}}\,.
\label{A1}
\end{align}
Here $\vec k_3 = \vec k_1-\vec k_2$, $n_1, n_2$ and $n_3$ are arbitrary integers; and $m_1, m_2$ and $m_3$ are arbitrary nonnegative 
real numbers.
If any $n_i$ is negative or equal to $0$, then we can assume that the corresponding $m_i=0$.

In several cases $f$ vanishes, namely, for arbitrary $n_1, n_2$ and $n_3$,
\begin{align}
f(0,0,0; n_1, n_2, n_3) = 0 \,.\label{A2}
\end{align}
If two parameters are equal to $0$, e.g., $m_1=m_2=0$, then
\begin{align}
f(0,0,m_3; n_1, n_2, n_3) = 0, \; \mbox{\rm for $n_1\leq 0$ or $n_2\leq 0$}\,. \label{A3}
\end{align}
In order to solve the general case, we first assume that $n_1=n_2=n_3=1$. Then the master integral $f$ is
\begin{align}
&f(m_1,m_2,m_3) \equiv f(m_1,m_2,m_3; 1,1,1) =
\nonumber \\
&4\,\pi\,\bigg(\frac{1}{4\,\epsilon}+\frac{1-\gamma_E + \ln(4\,\pi)}{2} - \ln(m_1+m_2+m_3)  \bigg)\,. \label{A4}
\end{align}
If all $m_i$ are not equal to $0$, then for positive $n_1, n_2$, and $n_3$,
\begin{align}
&f(m_1, m_2,m_3, n_1+1, n_2+1, n_3+1) =
\frac{(-1)^{n_1+n_2+n_3}}{n_1!\,n_2!\,n_3!}\nonumber \\ &\times
\frac{\partial^{n_1}}{\partial(m_1^2)^{n_1}}\,\frac{\partial^{n_2}}{\partial(m_2^2)^{n_2}}\,\frac{\partial^{n_3}}{\partial(m_3^2)^{n_3}}\,f(m_1,m_2,m_3) 
\,.\label{A5}
\end{align}
If $m_1=0$, then we use a general formula valid for arbitrary $n_i$,
\begin{widetext}
\begin{align}
 f(0,m_2, m_3; n_1,n_2,n_3)
  = &\
  \frac{m_3^{2\,(d-n_1-n_2-n_3)}}{(4\,\pi)^{d-3}}\,
  \frac{\Gamma(d/2-n_1)\, \Gamma(n_1+n_2-d/2)\, \Gamma(n_1+n_3-d/2)\, \Gamma(n_1+n_2+n_3-d)}
       {\Gamma(2\,n_1+n_2+n_3-d)\, \Gamma(n_2)\, \Gamma(n_3)\, \Gamma(d/2)}\nonumber \\&\
\times       _2\!F_1(n_1+n_2+n_3-d,n_1+n_2-d/2,2\,n_1+n_2+n_3-d,1- m_2^2/m_3^2)\,. \label{A6}
\end{align}
In fact, when $m_1=0$, we can still use differentiation as in Eq. (\ref{A5}), but later we have to separate out $1/m_1^i$
terms and the remainder coincides with the general formula in Eq. (\ref{A6}).
Two non obvious examples are
\begin{align}
f(0,m_2,m_3,-1,1,1)=&\  -4\,\pi\,m_2\,m_3\,(m_2^2 + m_3^2)\,, \label{A7}\\
f(0,m_2,m_3,-2,1,1)=&\  \frac{4\,\pi}{3}\,m_2\,m_3\,(3\,m_2^2 + m_3^2)\,(m_2^2 + 3\,m_3^2)\,. \label{A8}
\end{align}

\section{Matrix elements for $S$ states}
\label{APPB}
The following matrix elements contain $1/\epsilon$ singularity
\begin{align}
\bigg\langle\bigg[\frac{1}{r^4}\bigg]_\epsilon\bigg\rangle \equiv &\ \bigg\langle\bigg( \vec \nabla \bigg[\frac{1}{r}\bigg]_\epsilon\bigg)^2\bigg\rangle
=  \bigg\langle\frac{1}{r^4}\bigg\rangle + \langle \pi\,\delta^d(r)\rangle\bigg(-\frac{2}{\epsilon} + 8\bigg),
\label{B1}\\
\bigg\langle\bigg[\frac{1}{r^3}\bigg]_\epsilon\bigg\rangle \equiv &\ \bigg\langle\bigg[\frac{1}{r}\bigg]_\epsilon^3\bigg\rangle
\hspace{6ex} = \bigg\langle\frac{1}{r^3}\bigg\rangle + \langle \pi\,\delta^d(r)\rangle\bigg(\frac{1}{\epsilon} + 2\bigg)
\,.
\label{B2}
\end{align}
Here
\begin{align}
 \bigg\langle\frac{1}{r^3}\bigg\rangle  =&\ \lim_{a\rightarrow0}\int_a^\infty \frac{d r}{r} \, f(r) + f(0)\,(\gamma+\ln a)
=
\frac{4}{n^3}\bigg(\frac12 - \frac{1}{2\,n} - \gamma - \Psi(n) + \ln\frac{n}{2}  \bigg),
 \\
 \bigg\langle\frac{1}{r^4}\bigg\rangle  =&\ \lim_{a\rightarrow0}\int_a^\infty \frac{d r}{r^2} \, f(r) -\frac{f(0)}{a}+ f'(0)\,(\gamma+\ln a)
=  \frac{8}{n^3}\bigg(-\frac53 + \frac{1}{2\,n} + \frac{1}{6\,n^2} + \gamma + \Psi(n) - \ln\frac{n}{2} \bigg)\,,
 \\
 f(r) = &\ \int d\Omega\, \phi^*(\vec r)\,\phi(\vec r)
\end{align}
and
\begin{align}
\bigg\langle \pi\,\delta^d(r)\frac{1}{(E-H)'} \frac{p^4}{8}  \bigg\rangle =&\ \frac{1}{n^3}\bigg(-\frac32 - \frac{1}{n} + \frac{5}{4\,n^2} + \gamma + \Psi(n) - \ln\frac{n}{2} \bigg)
- \frac{1}{4\,\epsilon}\langle\pi\,\delta^d(r)\rangle
 \,, \label{B3}\\
\bigg\langle \pi\,\delta^d(r) \frac{1}{(E-H)'}  \pi\,\delta^d(r) \bigg\rangle =&\ \frac{1}{n^3}\bigg(- \frac12  - \frac{1}{n} + \gamma + \Psi(n) - \ln\frac{n}{2}  \bigg) - \frac{1}{4\,\epsilon}\langle\pi\,\delta^d(r)\rangle
\,, \label{B4}\\
\bigg \langle \pi\,\delta^d(r)\frac{1}{(E-H)'} \frac{1}{2}\,p^i\,\bigg[\frac{\delta^{ij}}{r} + \frac{r^i r^j}{r^3}\bigg]_\epsilon\,p^j  \bigg \rangle =&\
\frac{2}{n^3}\bigg(-\frac94 - \frac{1}{n} + \frac{5}{4\,n^2} + \gamma + \Psi(n) - \ln\frac{n}{2} \bigg) - \frac{1}{2\,\epsilon}\langle\pi\,\delta^d(r)\rangle \,. \label{B5}	
\end{align}
\end{widetext}
The regular matrix elements in a.u. are evaluated as
\begin{align}
E = &\ - \frac{1}{2\,n^2}\,, \label{B6}
\\
\bigg\langle \frac{1}{r}\bigg\rangle = &\ \frac{1}{n^2}, \label{B7}
\\
\bigg\langle \frac{1}{r^2}\bigg\rangle = &\ \frac{2}{n^3}, \label{B8}
\\
\langle \pi\,\delta^3(r)\rangle = &\ \frac{1}{n^3},  \label{B11}
\\
\big\langle \pi\,\nabla^2 \delta^d(r) \big\rangle =&\ \frac{2}{n^5}\,,  \label{B12}
\\
\big\langle \big\{ p^i , \big\{ p^i , \pi\,\delta^d(r)\big\}\!\big\} \big\rangle =&\ -\frac{2}{n^5}\,,  \label{B13}
\\
\bigg\langle p^i\,\bigg(\frac{\delta^{ij}}{r} + \frac{r^i r^j}{r^3}\bigg)p^j\bigg\rangle = &\ \frac{1}{n^3}\bigg(-\frac{2}{n}+4\bigg)\,,  \label{B14}
\\
\bigg\langle p^i\,\bigg(\frac{\delta^{ij}}{r^2} + \frac{r^i r^j}{r^4}\bigg)p^j\bigg\rangle = &\ \frac{1}{n^3}\bigg(-\frac{4}{3\,n^2}+\frac{16}{3}\bigg)\,. \label{B15}
\end{align}

\end{document}